\documentstyle[amssymb,aps,psfig,epsfig,graphics,twocolumn]{revtex}

\begin{document}
\title{Asymptotic solutions to the Gross-Pitaevskii gain equation: \\
growth of a Bose-Einstein condensate}
\author{P. D. Drummond and K. V. Kheruntsyan}
\address{{\it Department of Physics, The University of Queensland, }\\
Brisbane, {\it Queensland 4072, Australia}}
\date{July 27, 2000}
\maketitle

\begin{abstract}
We give an asymptotic analytic solution for the generic atom-laser system
with gain in a $D$-dimensional trap, and show that this has a
non-Thomas-Fermi behavior. The effect is due to Bose-enhanced condensate
growth, which creates a local density maximum and a corresponding outward
momentum component. In addition, the solution predicts amplified
center-of-mass oscillations, leading to enhanced center-of-mass temperature.%
\end{abstract}

\section{Introduction}

The description of Bose-Einstein condensate (BEC) growth \cite
{Ketterle,Kagan et al.,Stoof,Gardiner-Ballagh-97-98} has become important
due to the need to understand the physics of atom lasers \cite{Atom lasers}.
These recently developed devices which emit coherent wave-like beams of
atoms promise a new generation of precision measurements, applications in
nanotechnology and novel tests of fundamental concepts in quantum theory.
While it is possible to perform first-principles numerical simulations of
the relevant equations \cite{Drummond-Corney}, a great deal of physical
insight can be obtained from an analytic solution.

In this paper, we give an analytic asymptotic solution to the
Gross-Pitaevskii equation describing the early stages of condensate growth
in a trap. The physical insight we obtain from this is that a growing
non-equilibrium condensate has a non-uniform momentum distribution across
the condensed region. As a result, the observed density behaves as though
the trap frequency is increased, relative to the usual Thomas-Fermi behavior
\cite{Thomas-Fermi} of an equilibrium BEC. In addition, our solutions show
center-of-mass oscillations whose kinetic energy is amplified with the
condensate growth. This implies that, while BEC's formed through evaporative
cooling may have a low temperature for their internal degrees of freedom,
the temperature for the center-of-mass motion is higher, leading to
increased noise in the velocity and direction of atom laser beams.

\section{Gross-Pitaevskii gain model}

We start by considering a commonly used model of a one-component trapped
Bose-Einstein condensate - the Gross-Pitaevskii (GP) equation \cite{Gross &
Pitaevskii} modified by a linear gain term $g$ \cite{Walls}, of the form:
\begin{equation}
\frac{\partial \Psi ({\bf x},t)}{\partial t}=\left[ g-i\left( -\frac{\hbar }{
2m}\nabla ^{2}+V({\bf x})+U|\Psi |^{2}\right) \right] \Psi \,.
\label{GP-gain}
\end{equation}
Here $\Psi ({\bf x},t)$ is the mean-field amplitude (so that $|\Psi ({\bf x}%
,t)|^{2}$ is the particle number density), $m$ is the atomic mass, and $U$
is the effective interaction potential. In the treatment of $D=1,2$ or $3$
space dimensions, $U$ is given by $U=4\pi \hbar aL^{D-3}/m$, where $a$ is
the scattering length and $L$ is the confinement length. The potential term $%
V({\bf x})$ is due to an optical or magnetic trap, which we assume is
harmonic. In the simplest rotationally symmetric case, the trap potential is
given by $V({\bf x})=m\omega ^{2}|{\bf x}|^{2}/(2\hbar )$, where $\omega $
is the trap oscillation frequency.

It is helpful to understand the physical implications of Eq. (\ref{GP-gain})
by an analysis of the relevant time scales. In the present case, there are
three different relevant time-scales. These are: the time-scale for the
condensate growth, $t_{g}=1/g$, the trap oscillation time-scale, $t_{\omega
}=1/\omega $, and the ``healing'' time-scale, $t_{h}=1/(U|\Psi ({\bf x}%
_{0},t)|^{2})$, associated with the mean-field interaction potential at the
condensate center-of-mass ${\bf x}_{0}$.

Since current BEC's typically have a relatively high density, the healing
time is usually the smallest, which results in a Thomas-Fermi (TF) type
equilibrium condensate. If the gain time-scale is long enough, then one may
expect that even during growth, the condensate will adiabatically follow the
TF solution. This is commonly assumed in analyzing experimental data \cite
{Ketterle}. Another possibility is that the gain time-scale is shorter than
the trap oscillation period, in which case we should no longer expect
adiabatic TF like behavior.

\subsection{Asymptotically growing solution}

To show this distinction, we now proceed to give an analytic non-equilibrium
solution of the GP equation with gain. Firstly, we expand the field in terms
of the amplitude and phase,
\begin{equation}
\Psi ({\bf x},t)=A({\bf x},t)e^{-i\phi ({\bf x},t)}/\sqrt{U},
\end{equation}
and obtain the following coupled equations:
\begin{eqnarray}
\frac{\partial A}{\partial t} &=&gA+\frac{\hbar }{2m}A\nabla ^{2}\phi +\frac{%
\hbar }{m}\nabla \phi \cdot \nabla A.  \label{delA/delt} \\
\frac{\partial \phi }{\partial t} &=&A^{2}+\frac{m\omega ^{2}}{2\hbar }|{\bf %
x}|^{2}  \nonumber \\
&&+\frac{\hbar }{2m}\left( |\nabla \phi |^{2}-\frac{1}{A}\nabla ^{2}A\right)
,  \label{delphi/delt}
\end{eqnarray}

Next, we wish to investigate possible asymptotic solutions for long times --
i.e., steadily growing solutions, valid some time after initial nucleation
of the condensate, yet before any gain saturation has occurred. We consider
first the rotationally symmetric case, where $\nabla ^{2}=\partial
_{r}^{2}+[(D-1)/r]\partial _{r}$, with $r=|{\bf x}|$. Following the
construction successfully used recently in optical fiber environments \cite
{Harvey} -- and extending these to the current multi-dimensional case of a
trapped BEC -- we suppose that the amplitude has a self-similar behavior at
large time. The phase is assumed to depend on the atomic density at the
condensate center, and to have a uniform chirp giving a radially dependent
outward momentum. Thus, we seek a solution in the form of:
\begin{eqnarray}
A &=&\lambda (t)f(r/\lambda (t)). \\
\phi &=&\frac{\lambda ^{2}(t)}{2\tilde{g}\tau }-\frac{m\tilde{g}}{2\hbar }%
r^{2},
\end{eqnarray}
Here $\lambda (t)$ is a scaling function, while $\tau $ and $\tilde{g}$ are
unknown coefficients. Using this construction, we find, from Eqs. (\ref
{delA/delt}) and (\ref{delphi/delt}):
\begin{equation}
\frac{\partial \lambda (t)}{\partial t}\left( f-\frac{rf^{\prime }}{\lambda }%
\right) =g\lambda f-f^{\prime }\tilde{g}r-\frac{D}{2}\lambda f\tilde{g},
\label{Amplitude}
\end{equation}
\begin{eqnarray}
\frac{\partial \lambda (t)}{\partial t}\left( \frac{1}{\tilde{g}\tau }%
\lambda \right) &=&(\lambda f)^{2}+\frac{\hbar \tilde{\omega}^{2}}{2m}r^{2}
\nonumber \\
&&-\frac{\hbar }{2m}\left( \frac{f^{\prime \prime }}{f\lambda ^{2}}+\frac{D-1%
}{f\lambda r}f^{\prime }\right) ,  \label{phase}
\end{eqnarray}
where we have defined $\tilde{\omega}^{2}\equiv \omega ^{2}+\tilde{g}^{2}$.

From the first (amplitude) equation, the solution requires the twin
conditions that:
\begin{equation}
\frac{\partial \lambda (t)}{\partial t}=\left( g-\frac{D\tilde{g}}{2}\right)
\lambda =\tilde{g}\lambda .
\end{equation}
This implies that the scaling function $\lambda $ grows exponentially with
time, having a solution of $\lambda (t)=\exp (\tilde{g}t)$, where we can
immediately solve for the growth rate, since clearly $\tilde{g}=2g/(D+2)$.
The physical interpretation of this equation is rather straightforward; the
growth of the amplitude at any radial point is reduced below the single-mode
growth rate $g$, due to an outward flow of atoms, which transfers part of
the increased density to a larger radius. This effect increases with
dimensionality of the space.

The phase equation can now be simplified using the fact that the last two
terms involving derivatives have terms in $1/\lambda $, and hence become
exponentially smaller than the earlier terms, at long times, as the healing
time becomes smaller. This immediately gives the following solution for $f$:
\begin{equation}
f=\left[
\begin{array}{l}
\sqrt{(1-|{\bf x}|^{2}/R(t)^{2})/\tau },\;\text{for}\;\;|{\bf x}|<R(t), \\
\\
0,\;\text{for\ \ }\,|{\bf x}|>R(t),
\end{array}
\right.
\end{equation}
where $R(t)$ is the maximum radius given by:
\begin{equation}
R(t)=e^{\tilde{g}t}\sqrt{2\hbar /(m\tau \tilde{\omega}^{2})}.  \label{R(t)}
\end{equation}

The remaining unknown constant $\tau $ is obtained by evaluating the
integral $N(t)=\int |\Psi ({\bf x},t)|^{2}d^{D}{\bf x}$ for the total number
of particles:
\begin{equation}
N(t)=e^{\tilde{g}(D+2)t}\left( \frac{2\hbar }{m\tilde{\omega}^{2}}\right)
^{D/2}\frac{2C_{D}\,\tau ^{-(D+2)/2}}{UD(D+2)}.  \label{N(t)}
\end{equation}
Here $C_{D}$ is a dimension-dependent constant, with $C_{1}=2$, $C_{2}=2\pi $%
, and $C_{3}=4\pi $. This gives the expected result that the overall atom
number grows exponentially with a gain constant of $2g=\tilde{g}(D+2)$; that
is, $N(t)=N(0)\exp (2gt)$. The reason for this apparent difference in gain
constants is that the relatively slower growth in density at the center of
the condensate is exactly compensated for by the increase in condensate
radius with time.

The constant $\tau $ is solved in terms of $N(0)$, so that:
\begin{equation}
\tau =\left[ \left( \frac{2\hbar }{m}\right) ^{D/2}\frac{2C_{D}}{UD(D+2)N(0)%
\tilde{\omega}^{D}}\right] ^{2/(D+2)}.  \label{alpha}
\end{equation}
This coincides with the healing time for a condensate with atom number $N(0)$%
, in TF equilibrium -- but with an effective trap oscillation frequency $%
\tilde{\omega}$ instead of $\omega $.

The final result for the particle number density $|\Psi ({\bf x}%
,t)|^{2}=\lambda ^{2}f^{2}/U$ is:
\begin{equation}
|\Psi ({\bf x},t)|^{2}=\frac{e^{2\tilde{g}t}}{\tau U}\left( 1-\frac{|{\bf x}%
|^{2}}{R(t)^{2}}\right) .
\end{equation}

While this gives an asymptotic solution in terms of the initial atom number,
the solution needs to be compared with the usual TF solution to understand
the physical implications. In particular, we can notice an interesting
scaling behavior for the radius, as a function of the trap frequency. In the
present case, the radius scales as:
\begin{equation}
R\propto \tilde{\omega}^{-2/(D+2)}=(\omega ^{2}+\tilde{g}^{2})^{-1/(D+2)}.
\end{equation}

In equilibrium TF solutions, there is a similar behavior, except that there
is no outward momentum term. As a consequence, the TF radius ($R_{TF}\propto
\omega ^{-2/(D+2)}$) is always larger at a given total atom number, and the
density is lower at the center, than in these non-equilibrium solutions with
gain present. The physical reason for this is simply the rapid Bose-enhanced
increase in number density (and hence pressure) at the condensate center
during the nucleation process. This effect becomes appreciable when $%
g>>\omega $, or $t_{g}<<1/\omega $, so that the time-scale for condensate
growth is faster than the trap oscillation period.

\subsection{Numerical results}

We now turn to comparisons between the analytic asymptotic solution and
exact numerical results. In Fig. 1 we plot the r.m.s. radius $\bar{r}(t)$ of
a growing BEC versus $t$. The result found from the above asymptotic
solution in $3D$ is simply given by $\bar{r}(t)=\sqrt{3/7}R(t)$. This is
represented by the dashed line, for $g=10\omega $ and $N(0)=10$. For
comparison, the TF r.m.s. radius for the same values of $N(t)$ is
represented by the dotted line. The full lines correspond to the results of
direct numerical simulations of the GP equation with gain, with initial
Gaussian wave functions of different widths. The values of other parameters
are chosen to correspond to a $^{87}$Rb BEC ($m=1.44$ kg, $U_{3}=5\times
10^{-17}$ m$^{3}$/s) in a trap with $\omega /2\pi =100$ Hz. As one can see
from the graph, despite the initial differences, the mean radii approach the
same asymptotic value of $\sim \sqrt{3/7}R(t)$, which is different to the TF
result for the same total number of particles.

The outward momentum density $p=|{\bf p}|\ $(where ${\bf p}=-i\hbar \lbrack
\Psi ^{\ast }({\bf \nabla }\Psi )-H.c.]/2$) is plotted in Fig. 2, showing
that in a growing BEC the flow of atoms from the trap centre increases
initially with the distance, reaches a maximum, and vanishes at $r\gtrsim R$.

\section{Asymmetric case and center-of-mass oscillations}

The above rotationally symmetric results can easily be generalized to
asymmetric cases, where the trap oscillation frequencies $\omega _{i}$ in
each space direction are different. In this case, the trap potential term in
Eqs. (\ref{GP-gain}) and (\ref{delphi/delt}) is replaced by
\begin{equation}
V({\bf x})=\frac{m}{2\hbar }\sum\nolimits_{i=1}^{D}\omega _{i}^{2}x_{i}^{2}.
\end{equation}

Assuming again that the amplitude has a self-similar behavior at long times,
one can use the previous construction. In addition, we find that a further
more general solution can be obtained. Following the approach of Kohn's
theorem \cite{Kohn}, generalized to include gain, we allow for
center-of-mass oscillation of the growing condensate, independent of the
inter-particle interactions.

Thus, we seek a solution in the form of
\begin{eqnarray*}
A &=&\lambda (t)f\left( \Delta {\bf x}(t)/\lambda (t)\right) , \\
\phi &=&\phi _{0}(t)+\frac{\lambda (t)^{2}}{2\tilde{g}\tau }-\frac{m}{\hbar }%
\left[ \frac{\tilde{g}\Delta {\bf x}(t)^{2}}{2}+\Delta {\bf x}(t)\,\cdot
{\bf \dot{x}}_{0}(t)\right] ,
\end{eqnarray*}
where $\Delta {\bf x}(t)\equiv {\bf x}-{\bf x}_{0}(t)$, ${\bf x}_{0}(t)$ is
the center-of-mass, and the dot stands for the time derivative.

Using the same procedure as before, we first find -- from the amplitude
equation -- that the function $\lambda (t)$ is given by
\begin{equation}
\lambda (t)=\exp (\tilde{g}t),
\end{equation}
where $\tilde{g}=2g/(D+2)$, as previously.

To treat the phase equation, in which we neglect the last term $\propto
1/\lambda $ that becomes exponentially small at long times, we assume that
\[
f^{2}=F_{1}+F_{2}(\Delta {\bf x)}^{2},
\]
where the functions $F_{1}$ and $F_{2}$ are to be found by equating the
terms in powers of $\Delta x_{i}$. From the terms in $\Delta x_{i}$, we find
that each component of ${\bf x}_{0}(t)$ satisfies equation $\ddot{x}%
_{0,i}+\omega _{i}x_{0,i}=0$, so that the condensate center-of-mass
oscillates according to
\begin{equation}
x_{0,i}(t)=a_{0,i}\cos (\omega _{i}t+\delta _{0,i}),
\end{equation}
where $a_{0,i}$ and $\delta _{0,i}$ are the initial amplitude and phase.

By equating the terms in $(\Delta x_{i}{\bf )}^{2}$ and the zeroth-order
terms, respectively, we obtain an equation for $\phi _{0}(t)$,
\[
\dot{\phi}_{0}+\frac{m}{2h}\sum\nolimits_{i=1}^{D}\left[ (\dot{x}%
_{0,i})^{2}-\omega _{i}^{2}x_{0,i}\right] =0,
\]
as well as solutions to $F_{1}$ and $F_{2}$:
\[
F_{1}=1/\tau ,
\]
\[
F_{2}=-\frac{m}{2\hbar \lambda (t)}\sum\nolimits_{i=1}^{D}\frac{(\omega
_{i}^{2}+\tilde{g}^{2})\Delta x_{i}(t)}{(\Delta {\bf x}(t))^{2}}.
\]

Combining these together and using the solutions for the center-of-mass
coordinates, we finally obtain that the solution for the function $\phi
_{0}(t)$ is given by
\begin{equation}
\phi _{0}(t)=-(m/2\hbar )({\bf x}_{0}(t)\,\cdot {\bf \partial \dot{x}}%
_{0}(t)),
\end{equation}
while the solution for $f$ is
\begin{equation}
f=\frac{1}{\sqrt{\tau }}\left( 1-\sum\nolimits_{i=1}^{D}\frac{%
(x_{i}-x_{0,i}(t))^{2}}{R_{i}(t)^{2}}\right) ^{1/2},
\end{equation}
in the region of space where the expression in large brackets is positive,
and $f=0$ outside that region.

The maximum radius $R_{i}(t)$ in each space direction is
\begin{equation}
R_{i}(t)=e^{\tilde{g}t}\sqrt{2\hbar /(m\tau \tilde{\omega}_{i}^{2})},
\end{equation}
where we have introduced $\tilde{\omega}_{i}^{2}\equiv \omega _{i}^{2}+%
\tilde{g}^{2}$. In addition, the constant $\tau $ is solved, as previously,
in terms of $N(0)$ by evaluating the integral for the total number of
particles $N(t)=N(0)\exp (2gt)$. The resulting expression for $\tau $ is
given by the same equation as before, Eq. (\ref{alpha}), except that $\tilde{%
\omega}^{D}$ is replaced by $\prod_{j=1}^{D}\tilde{\omega}_{j}$. This again
corresponds to the healing time in a TF condensate, with $N(0)$ particles
and effective trap oscillation frequencies $\tilde{\omega}_{i}$.

The final result for the particle number density is now:
\begin{equation}
|\Psi ({\bf x},t)|^{2}=\frac{e^{2\tilde{g}t}}{\tau U}\left( 1-\sum_{i=1}^{D}%
\frac{\left( x_{i}-x_{0,i}(t)\right) ^{2}}{R_{i}(t)^{2}}\right) .
\end{equation}
An example showing growth of the BEC, while the condensate center-of-mass
oscillates, is shown in Fig. 3.

\section{Summary}

The physical interpretation of these results is that the asymptotic solution
with gain has a density distribution that is similar to the TF solution,
except that the trap oscillation effective frequency is increased, with $%
\omega _{i}^{2}$ $\rightarrow $ $\tilde{\omega}_{i}^{2}=\omega _{i}^{2}+%
\tilde{g}^{2}$. For a given gain constant, this has the strongest effect for
weakly trapped (low-frequency) directions, in an asymmetric trap. In
addition, the solution gives an outward momentum component due to bosonic
stimulation effect which is stronger in the center of the trap.

In addition, the results obtained show that Bose stimulation can occur to
moving condensates as easily as to stationary ones. In cases like this, the
amplitude of the center-of-mass oscillation does not change during the
condensate growth, while the total mass of the condensate increases. This
means that the condensate center-of-mass kinetic energy, $%
E_{K}=N(t)m\sum_{i}\omega _{i}^{2}a_{0,i}^{2}$, can grow exponentially to
large values -- even in the absence of technical noise. While the present
model of gain is simplified, similar types of center-of-mass motion are
found in first-principles simulations of condensate formation \cite
{Drummond-Corney}. This suggests that, while BEC's formed through stimulated
emission may have a low temperature relative to the center-of-mass, the
center-of-mass motion may itself have a higher temperature. Current
experimental measurements of single atom velocity distributions \cite{Wieman}
appear insensitive to the center-of-mass temperature of the condensate. The
effect is analogous to the increased uncertainty in the frequency and
pointing stability of a multi-mode optical laser, compared to a single-mode
laser.

We emphasize that the asymptotic solutions given here are only applicable
for condensate nuclei that have already formed, as the spontaneous emission
noise is omitted. The gain medium is assumed to be unsaturated and to
equilibrate rapidly, giving a uniform gain constant across the growing
condensate. In addition, one can expect damping to occur due to interactions
with non-condensed atoms \cite{Griffin}, which are not treated here. These
interactions, however, can only equilibrate the temperatures of condensed
and non-condensed center-of-mass motion, rather than providing additional
cooling of the condensate center-of-mass motion. Other damping mechanisms
that will intrinsically be present in a BEC are two-body losses, which
effectively change $g$ to $g-\Gamma |\Psi |^{2}$, where $\Gamma $ is the
two-body loss rate. While these affect the form of the asymptotic solution,
two-body (or higher-order) losses do not change the amplitude of the
condensate center-of-mass oscillation

In summary, we have found an asymptotic solution to the Gross-Pitaevskii
equation with gain, which has the advantage of yielding an explicit analytic
result of great physical transparency. The solution shows that the
non-equilibrium behavior of a growing Bose-Einstein condensate generally
includes an outward momentum component and spatial oscillations. As a
result, we suggest that the state of a trapped BEC is not a canonical
ensemble, and should be characterized by at least two distinct temperatures:
one for the internal degrees of freedom, and one for the center-of-mass
motion.

The authors acknowledge the Australian Research Council for the support of
this work.

\bigskip

\begin{figure}[h]
\centerline{\hbox{
\epsfig{figure=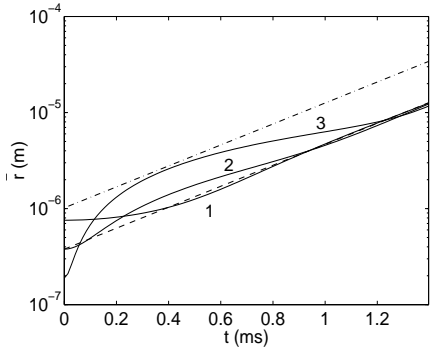,height=4.7 cm,width=5.7 cm}}}
\vspace{0.3 cm}
\caption{The r.m.s. radius $\bar{r}$ (in a logarithmic scale) of a growing
condensate versus $t$, for $g=10\omega $. The full lines (1), (2) and (3)
correspond, respectively, to initial Gaussian wave functions having r.m.s.
radii that are twice larger, equal, and twice smaller than the asymptotic
r.m.s. radius $\bar{r}(0)=\sqrt{3/7}R(0)$. The dashed line is the asymptotic
result, while the dotted line corresponds to the equilibrium (TF) solution.}
\end{figure}

\begin{figure}[h]
\centerline{\hbox{
\epsfig{figure=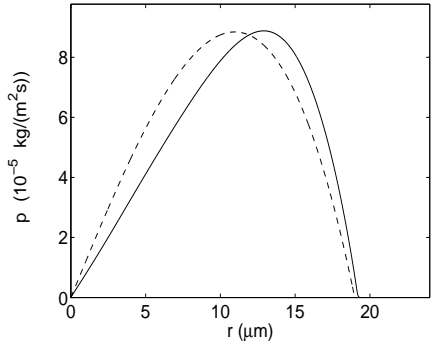,height=4.7 cm,width=5.7 cm}}}
\vspace{0.3 cm}
\caption{The momentum density $p\ $versus the radial distance from the centre
of the trap $r$. The full line corresponds to the resulting distribution of
case (1) in Fig. 1; the dashed line is the corresponding asymptotic solution.
}
\end{figure}

\begin{figure}[h]
\centerline{\hbox{
\epsfig{figure=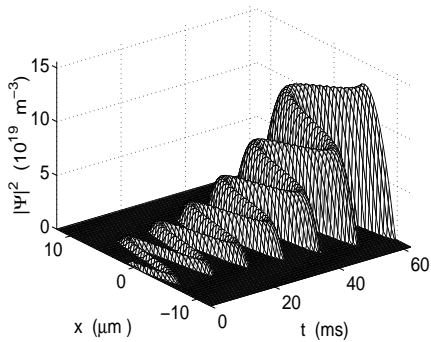,height=4.7 cm,width=5.7 cm}}}
\vspace{0.3 cm}
\caption{Growth of a BEC with a center-of-mass oscillation present. Shown is
the condensate density $|\Psi (x,t)|^{2}$, as found from the asymptotic
solution in a symmetric trap, with a center-of-mass oscillation in the $x$%
-direction and $N(0)=10$. The gain coefficient is chosen as $g=0.1\omega $,
while $a_{0,1}=4.86$ $\mu $m, and $\delta _{0,1}=\pi /2$. Other parameter
values are as previously.}
\end{figure}

\end{document}